\documentclass[final,5p,times,twocolumn]{elsarticle}

\usepackage{graphicx}

\usepackage{amssymb}

\usepackage{rotating}
\journal{NIMB}

\begin{document}

\begin{frontmatter}



\title{Effective electron recombination coefficient in ionospheric D-region during
 the relaxation regime after solar flare from February 18, 2011 }

%

\author[label1]{A.~Nina}
\author[label2]{V.~\v{C}ade\v{z}}
\author[label3]{D.~\v{S}uli\'{c}}
\ead{dsulic@ipb.ac.rs}
\author[label1]{V.~Sre\'{c}kovi\'{c}}
\author[label4]{V.~\v{Z}igman}

\address[label1]{Institute of
Physics, University of Belgrade, P.O. Box 57,  Belgrade, Serbia}
\address[label2]{Astronomical Observatory, Volgina 7, 11060 Belgrade, Serbia}
\address[label3]{Faculty of Ecology and Environmental Protection, Union - Nikola Tesla University, Cara Du\v{s}ana 62, 11000 Belgrade, Serbia}
\address[label4]{University of Nova Gorica,Vipavska 13,Rožna Dolina, SI-5000 Nova
Gorica, Slovenia}

\begin{abstract}
In this paper, we present a model for determination of a weakly time
dependent effective recombination coefficient for the perturbed
terrestrial ionospheric D-region plasma. We study consequences of a
class M1.0 X-ray solar flare, recorded by GOES-15 satellite on
February 18, 2011 between 14:00 UT and 14:15 UT, by analyzing the
amplitude and phase real time variations of very low frequency (VLF)
radio waves emitted by transmitter DHO (located in Germany) at
frequency 23.4 kHz and recorded by the AWESOME receiver in Belgrade
(Serbia). Our analysis is limited to ionospheric perturbations
localized at altitudes around 70 km where the dominant electron gain
and electron loss processes are the photo-ionization and
recombination respectively.
\end{abstract}

\end{frontmatter}

\section{Introduction}
\label{intr} The ionosphere, having characteristics of plasma, is
very sensitive to electromagnetic disturbances whose intensity and
number vary with solar activity. These disturbances cause numerous
complicated physical, chemical and dynamical phenomena in the lower
ionosphere and may directly affect human activities, especially in
the telecommunications industry. Besides a pure scientific interest
to study the influence of solar activity on the terrestrial
atmosphere, the understanding and predicting the resulting turbulent
regions of the ionosphere has important applications for radio
communications, military operations in remote locations, planned
networks of mobile communications satellites, high-precision
applications of global navigation satellite systems, etc.

The VLF radio signal propagation properties are determined by the
wave attenuation and reflection which depend on the electron
concentration and collision frequency. The electron concentration
calculations in empirical and semi-empirical models are based mainly
on measurements using radio-wave propagation methods and sounding
rockets.

Due to a varying complex structure of the atmosphere and because of
numerous influences coming from the Earth and outer space,
experimental simulations of processes in the ionosphere are very
difficult. For this reason, theoretical models are of a basic
importance in providing us with information on ionospheric
parameters, with the effective recombination coefficient
$\alpha_{\rm eff}$ being one of them. This coefficient plays an
important role in the ionization balance and in kinetics of
underlying chemical reactions and from its values one can estimate
the structure of the lower ionosphere.

In the Earth ionosphere, the electron concentration depends on
photo-ionization and recombination processes. At heights around 70
km, the dominant recombination processes are the electron-ion,
ion-ion and three-body recombination characterized by a
common effective recombination coefficient
$\alpha_{\rm eff}$ \cite{zig07}.

In the case of unperturbed, quasi-stationary ionosphere, the
electron gain and electron loss processes are in equilibrium, i.e.
mutually balanced.

During a solar flare, the electron concentration is raised due to
increased photo-ionization and this time period can be named as
photo-ionization regime. After a flare maximum when the rate of
recombination becomes larger than the photo-ionization rate, the
electron concentration starts to decrease and the ionosphere enters
a recombination regime.

In this paper, we calculate the time dependence of effective
recombination coefficient $\alpha_{\rm eff}$ for different altitudes
in the ionospheric D-region for the tail of the recombination
regime. The considered class M1.0 X-ray solar flare was recorded by
GOES-15 satellite on February 18, 2011 between 14:00 and 14:15 UT.
Our analysis uses a discrete set of data on amplitude and phase
variations of the 23.4 kHz VLF signal emitted by the transmitter DHO
in Germany and registered by the AWESOME \footnote{Atmospheric
Weather Electromagnetic System for Observation Modeling and
Education.} receiver system \cite{sch08} in Belgrade\footnote{A part
of Stanford/AWESOME Collaboration for Global VLF Research.}, Serbia
with the sampling period of 0.02 s. These experimental data were
treated by the LWPC (Long-Wave Propagation Capability) computer
program \cite{fer98} for modeling the ionosphere which yielded
discrete sets of data on the wave reflection heights $H^\prime(t)$
and sharpness $\beta(t)$ the basic parameters of Wait$^\prime$s
model of the ionosphere \cite{wai64} needed to compute the
distribution profiles of electron concentration $N(t,h)$ and the
effective recombination coefficient $\alpha_{\rm eff}(t,h)$ where
$h$ is the location altitude.

\section{Basic theory}
\label{theo}

The electron concentration $N(t,h)$ at altitude $h$ in the D-region
is time dependent and is taken to obey the following differential
equation \cite{app53}:

\begin{equation}
{dN(t,h) \over dt} = q(t,h)-\alpha_{\rm eff}(t,h) N^2(t,h),
\label{e1}
\end{equation}
where the concentrations of negative ions and their time derivatives
are comparatively small at altitudes of about 70 km \cite{ris69} so
that their contributions are not taken into account.

According to the Chapman$^\prime$s theory \cite{cha31} the rate of
electron production $q(t,h)$ can be expressed as:
\begin{equation}
q(t,h) = Q_1(t,h)I(t,h),
 \label{e2}
\end{equation}
where $I(t,h)$ is the X-ray radiation flux intensity and $Q_1(t,h)$
is a proportionality coefficient whose explicit expression turns out
not to be required in our computations.

The X-ray radiation flux intensity $I(t,h)$ at some level $h$ is
related to its value $I_\infty$ (measured by the satellite GOES-15)
through the following relation:

\begin{equation}
I(t,h) = Q_2(t,h)I_\infty(t),
\label{e3}
\end{equation}
where $Q_2(t,h)$, like $Q_1(t,h)$ in Eq.~(\ref{e2}), is a
proportionality coefficient whose explicit expression will not be
needed in our further treatment.

From Eqs (\ref{e1})-(\ref{e3}) we can write:

\begin{equation}
\frac{dN(t,h)}{dt} = K(t,h)I_\infty(t)-\alpha_{\rm eff}(t,h)
N^2(t,h), \label{e4}
\end{equation}
with the coefficient $K(t,h)$ given by:

\begin{equation}
K(t,h) \equiv Q_1(t,h)Q_2(t,h). \label{e5}
\end{equation}

After a sufficient time interval following the intensity maximum
(the shaded domain in Fig.~\ref{f1}), the coefficients $K(t,h)$ and
$\alpha_{\rm eff}(t,h)$ become weakly time dependent and we will
consider them approximately constant within some finite time
interval $\Delta t$:

\[
K(t-\Delta t,h)\approx K(t,h)\equiv \overline{K}(t,h),
\]
\[\alpha_{\rm eff}(t-\Delta t,h)\approx \alpha_{\rm eff}(t,h)
\equiv\overline{\alpha}_{\rm eff}(t,h).
\]

Eq.~(\ref{e4}) can now be applied to the interval end points
 $t_{\rm 1}=t-\Delta t$ and $t_{\rm 2}=t$ which yields the following set
 of two algebraic equations for unknown quantities $\overline{K}(t,h)$ and
 $\overline{\alpha}_{\rm eff}(t,h)$:

\begin{equation}
\begin{array}{l}
\displaystyle\left.\frac{dN}{dt}\right\vert_{t-\Delta t,h}=\overline{K}(t,h)I_\infty(t-\Delta t)\\[.2cm]
\hspace{1.8cm}-\overline{\alpha}_{\rm
eff}(t,h) N^2(t-\Delta t,h)\\[.2cm]
\displaystyle\left.\frac{dN}{dt}\right\vert_{t,h}=\overline{K}(t,h)I_\infty(t)
-\overline{\alpha}_{\rm eff}(t,h) N^2(t,h)
\end{array}
\label{e6}
\end{equation}
which finally gives the expression for numerical computation of the
weakly time dependent effective recombination coefficient
$\overline{\alpha}_{\rm eff}(t,h)$:

\begin{equation}
\begin{array}{ll}
\displaystyle \overline{\alpha}_{\rm eff}(t,h)=
\frac{I_\infty(t-\Delta t)}{ {\cal A}(t,\Delta t,h)}
\left.\frac{dN}{dt}\right\vert_{t,h} -\frac{I_\infty(t)}{{\cal
A}(t,\Delta t,h)}\left.\frac{dN}{dt}\right\vert_{t-\Delta t,h}
\end{array}
\label{e7}
\end{equation}
where:
\begin{equation}
{\cal A}(t,\Delta t,h) \equiv N^2(t-\Delta
t,h)I_\infty(t)-N^2(t,h)I_\infty(t-\Delta t).
\end{equation}

All quantities on the right-hand-side in Eq.~(\ref{e7}) have values
that can easily be obtained from both the recorded signal properties
and Wait$^\prime$s model of the ionosphere \cite{wai64}. In
computations that follow, we take $\Delta t =1$s.

\section{Experimental data and calculation procedure}
\label{exsp}

\begin{figure}[h]
\begin{center}
\includegraphics[width=2.5in]{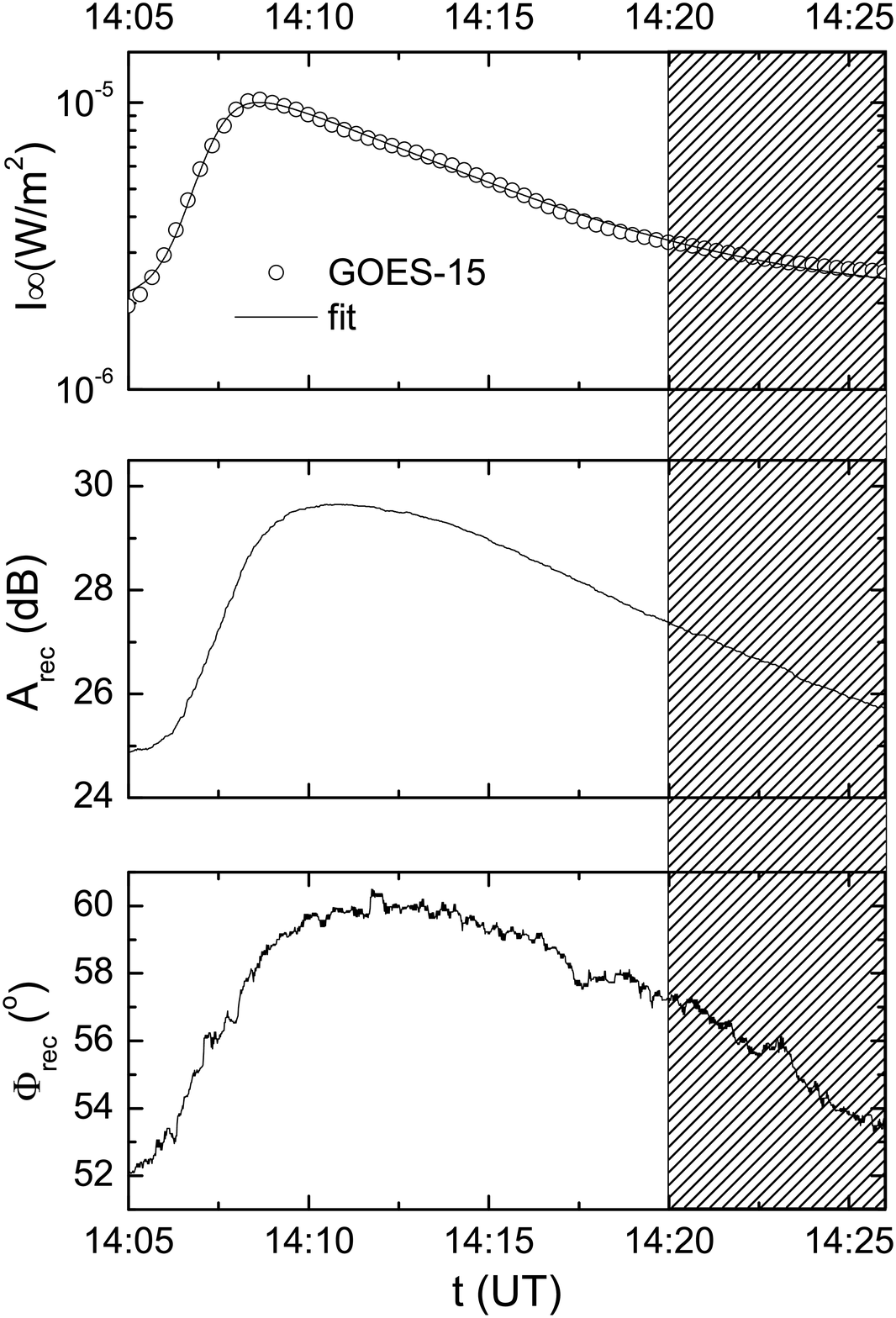}
\caption{Solar flux registered by GOES-15 satellite and perturbed
amplitude and phase of signal emitted from transmitter DHO (Germany)
and recorded on AWESOME receiver in Belgrade (Serbia) during
observed flares. } \label{f1}
\end{center}
\end{figure}

Within the time interval 14:00 UT - 14:15 UT, February 18, 2011, a
solar flux increase was registered in the range 0.1-0.8 nm by the
GOES-15 satellite whose maximal value corresponded to the level
typical of a class M1.0 solar X flare (Fig.~\ref{f1}). This rise on
the radiation flux altered the rate of photo-ionization and electron
recombination process and, consequently, it changed the electron
concentration. As a result, the recorded VLF wave exhibits a time
varying amplitude and phase as shown in Fig.~\ref{f1}.

To obtain the effective recombination coefficient $\alpha_{\rm
eff}(t,h)$ profiles given by Eqs.~(\ref{e7})-(\ref{e8}) we need the
explicit expression for the time dependence of the electron
concentration profile $N$ and its time derivative $dN/dt$ as well as
the incoming solar radiation flux as recorded by the GOES-15
satellite.

\begin{figure}
\begin{center}
\includegraphics[width=2.5in]{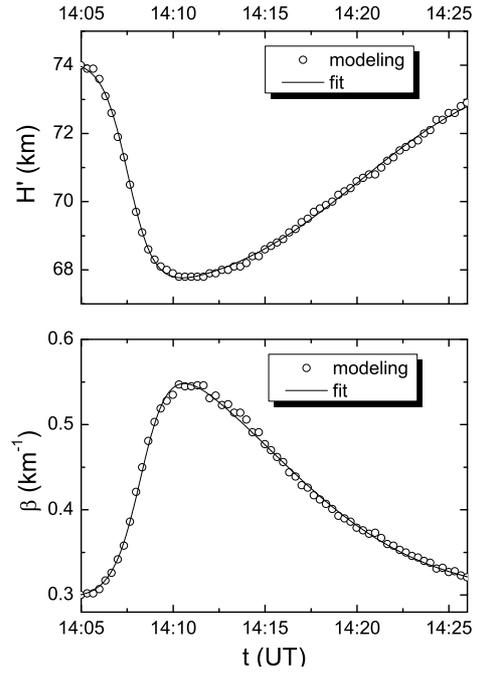}
 \caption{The reflection height $H'$ and sharpness $\beta$ obtained by comparison LWPC simulation and
recorded amplitude and phase values.} \label{f2}
\end{center}
\end{figure}

The electron concentration $N(t,h)$ is computed by the following
relation \cite{wai64}:
\begin{equation}
N(t,h) = 1.43\cdot10^{13}e^{-\beta(t)H'(t)}e^{(\beta(t)-0.15)h},
\label{e8}
\end{equation}
where  $H'(t)$ and $\beta(t)$ are defined as the wave reflection
height and sharpness respectively. Quantities $H'(t)$ and $\beta(t)$
are determined by the LWPC computer program \cite{fer98} using the
recorded discrete time-set of values for the amplitude and phase of
the considered VLF 23.4 kHz VLF signal (\cite{gru08,sul10}). Time
dependencies of these two parameters are shown in Fig.~\ref{f2}
where it can be seen that the VLF wave reflection height $H'(t)$
initially decreases in time and reaches a minimum at about two
minutes after the solar flux maximum (Fig.~\ref{f2}). The sharpness
$\beta(t)$, characterizing the gradient of the electron
concentration with height, has a reversed time distribution, i.e. it
first increases and attains a maximum at the same time when $H'(t)$
has its minimum.

 As the recorded quantities are discrete in
time, the input radiation flux intensity $I_\infty(t)$ and the
derived values for $\beta(t)$ and $H'(t)$ are not smooth functions
of time $t$ and we fitted them by a continuous model function
implemented in the Origin-8 program, also found in the NAG
(Numerical Algorithms Group) library:
\begin{equation}
y =
y_0+\frac{A}{1+e^{-\frac{x-x_c+w_1/2}{w_2}}}\left(1-\frac{1}{1+e^{-\frac{x-x_c-w_1/2}{w_3}}}
\right), \label{e9}
\end{equation}
where $x$ is time $t$ in seconds starting from 14:04:00 UT and $y$
stands for $I_\infty(t)$, $\beta(t)$, and $H'(t)$. The numerical
fitting constants $y_0$, $A$, $x_c$, $w_1$, $w_2$ and $w_3$ related
to the considered physical quantities $y$ in Eq.~(\ref{e9}) were
obtained as the program outputs and are presented in Table 1.

Thus derived continuous time dependent function $N(t,h)$ yields a
smooth time derivative $dN/dt$ and the effective recombination
coefficient $\alpha_{\rm eff}(t,h)$ according to
 Eqs.~(\ref{e7})-(\ref{e8}).

\section{Results and discussion}
\label{res}

\begin{figure}
\begin{center}
\includegraphics[width=2.5in]{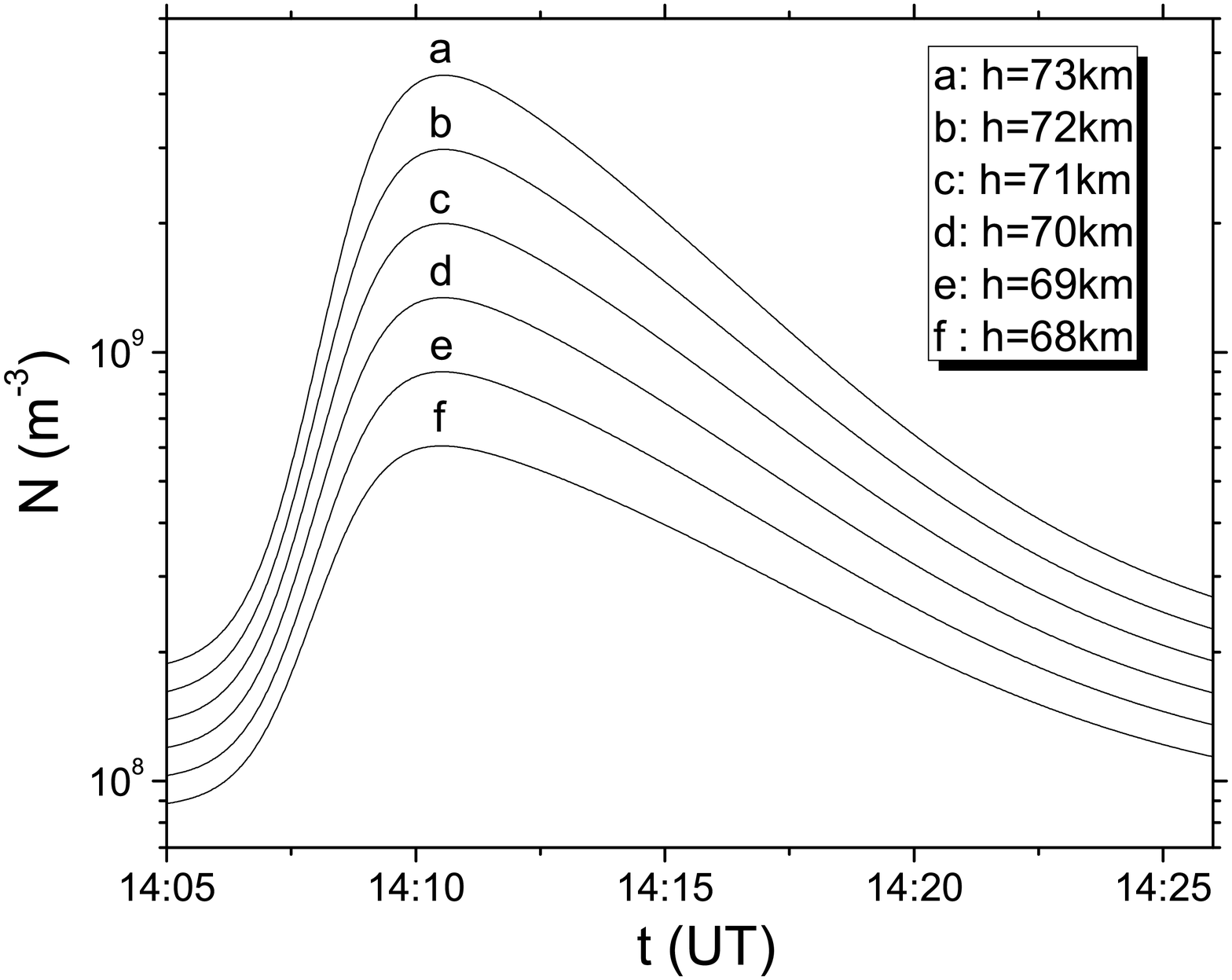}
\caption{The time distribution of electron concentration during
photo-ionization and recombination regimes.} \label{f3}
\end{center}
\end{figure}

\begin{figure}
\begin{center}
\includegraphics[width=2.5in]{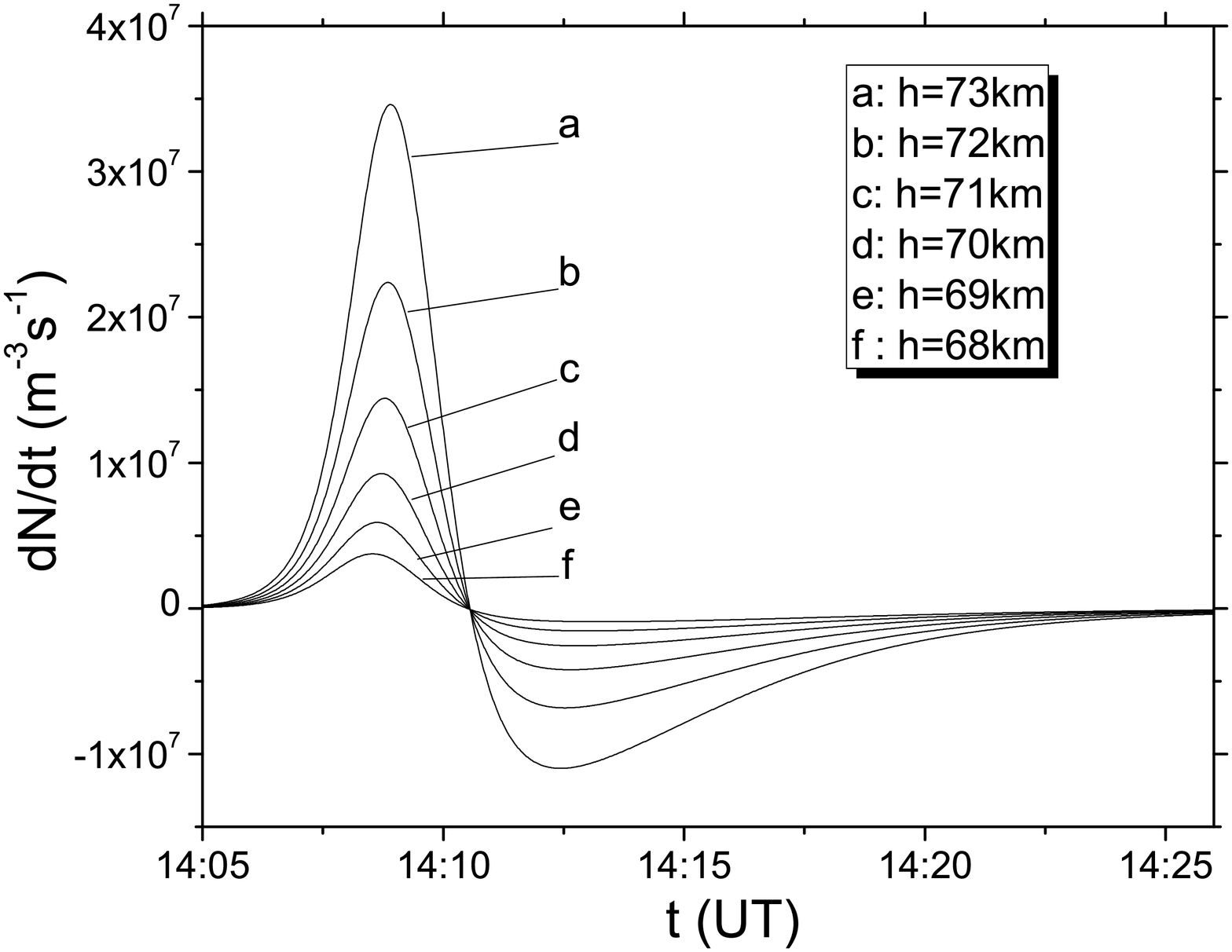}
\caption{The time distribution of the time derivative of electron
concentration during photo-ionization and recombination regimes.}
\label{f4}
\end{center}
\end{figure}

The electron concentrations at various heights and their time
derivatives obtained from Eq.~(\ref{e8})  are shown in Figs \ref{f3}
and \ref{f4} respectively. We can see that the considered solar
flare causes larger increases in electron concentration at higher
altitudes. Comparing Fig.~\ref{f1} and Fig.~\ref{f3} it can be
noticed that the time distribution of the electron concentration
follows the time variation pattern of the registered solar flux on
GOES-15 satellite. Maximum of solar flux occurs at 11:24 UT which is
about two minutes before the time when the corresponding maximum
appears in the electron concentration curve. At considered altitudes
of around 70 km, the dominant electron gain and electron loss
processes are the photo-ionization and recombination respectively.
Within the time period before the electron concentration maximum,
the rate of the photo-ionization is larger then that of
recombination processes. Contrary, the domination of recombination
processes causes a decreasing electron concentration as seen in the
recombination regime in Fig.~\ref{f3}.

The rate of the electron concentration change is more pronounced at
higher altitudes. It is positive when photo-ionization processes
dominate the recombination processes. Contrary, the domination of
the recombination produces a time decrease of the electron
concentration and, consequently, its negative time derivative.

\begin{figure}
\begin{center}
\includegraphics[width=2.5in]{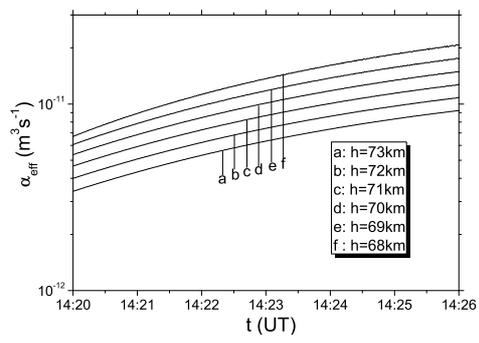}
\caption{Effective electron recombination coefficient in ionospheric
D-region for the tail of
 the relaxation regime following solar flare on February 18, 2011.} \label{f5}
\end{center}
\end{figure}

In Fig.~\ref{f1}, the shaded time domain indicates a relatively slow
time variation of indicated quantities that allows us to use the
procedure for calculation effective recombination coefficient
$\alpha_{\rm eff}(t,h)$ according to the described basic theory.
Time variations of $\alpha_{\rm eff}(t,h)$ are presented in
Fig.~\ref{f5} for different altitudes. Unlike the electron
concentration, the effective recombination coefficient has larger
values at smaller altitudes.

The increase of coefficient $\alpha_{\rm eff}(t,h)$ is in agreement
with \cite{zig07} where $\alpha_{\rm eff}(t,h)$ has larger values
for smaller maxima of flare radiation intensity.

\section{Conclusion}
\label{concl}

In this work, we present a method to  calculate the effective
recombination coefficient at altitudes around 70 km for the tail of
the recombination regime. It is shown that the effective
recombination coefficient depends on the solar flux intensity,
electron concentration and its time derivative. We present a method
of determining the electron concentration by comparing the recorded
VLF signal amplitude and phase changes with the corresponding values
obtained numerically using the LWPC computer program. In the case of
this flare, the effective recombination coefficient has values
within the order of magnitude $10^{-12}$ $\rm m^3\rm s^{-1}$ to
$10^{-11}$ $\rm m^3\rm s^{-1}$.

\section*{Acknowledgment}
The present work has been performed under the Ministry of Education
and Science of the Republic of Serbia (projects III 44002, 176002
and 176004), Slovenian Research Agency, contract P2-0056 and Joint
Bilateral project: BI$_-$SLO$_-$SR$/10-11-038$.

\begin{sidewaystable}
\begin{center}
\caption{Parameters used in Eq.~(\ref{e9}).}
\begin{tabular}{|c|c|c|c|c|c|c|}
\hline
 $y$&$y_0$&$A$&$x_c$&$w_1$&$w_2$&$w_3$\\
\hline
$I_\infty(t)$&2.07417E-6&1.97397E-5&196.87571&7.05136E-37&29.81728&282.37906\\
(W/m$^2)$&W/m$^2$&W/m$^2$&s&s&s&s\\
\hline
$\beta(t)$&0.29789&0.33295&0.59163&0.0051&5.05997E-4&0.00278\\
(km$^{-1})$&km$^{-1}$&km$^{-1}$&s&s&s&s\\
\hline
$H'(t)$&74.13507&-7.06355&590.67954&754.49827&41.86687&242.7825\\
(km)&km&km&s&s&s&s\\
\hline
\end{tabular}
\end{center}
\end{sidewaystable}

\end{document}